\documentclass[preprint,12pt]{elsarticle}

\usepackage[pdftex]{color}
\usepackage[font=footnotesize,labelfont=bf]{caption}
\usepackage[font=footnotesize,labelfont=bf]{subcaption}

\usepackage[margin=2.5cm]{geometry}
\usepackage{amssymb}

\usepackage{amsmath}
\usepackage{cleveref}

\def \wireAx  { \rm  Ax. }
\def \st  {  \rm  St. }
\def \ax  {  \rm Ax. }

\journal{Nuclear Physics B}
\makeatletter
\def\ps@pprintTitle{%
  \let\@oddhead\@empty
  \let\@evenhead\@empty
  \def\@oddfoot{\reset@font\hfil\thepage\hfil}
  \let\@evenfoot\@oddfoot
}
\makeatother

\begin{document}

\begin{frontmatter}



\title{GPU-accelerated  Interval Arithmetic  to solve the Apollonius Problem applied to a Stereo Drift Chamber}


\author[inst1]{Wilfrid da Silva}
\author[inst2]{Patrice Lebrun}
\author[inst3]{Jean-Claude Ang\'elique}
\author[inst1]{Luigi Del Buono}
\affiliation[inst1]{organization={LPNHE, Sorbonne Université, CNRS/IN2P3, Université Paris Cité, UMR 7585},
            city={Paris},
            country={France}}
\affiliation[inst2]{organization={Université de Lyon, Université Claude Bernard Lyon 1, CNRS/IN2P3, Institut de Physique des 2 Infinis de Lyon, UMR 5822},
            city={Villeurbanne},
            country={France}}    
\affiliation[inst3]{organization={Université de Caen Normandie, ENSICAEN, CNRS/IN2P3, LPC Caen,  UMR6534},
            city={Caen},
            country={France}}         
\begin{abstract}
We propose a new system of equations which identifies the helix 
common to all drift distance hits produced by a full stereo
drift chamber detector when a charged particle 
passes through this detector. The 
equation system 
is obtained using the Apollonius' problem as guideline which 
gives it a very simple form and a clear physics interpretability 
as the case of full axial  drift chamber detector. The proposed 
method is evaluated using drift  distance hits constructed from 
Monte Carlo-generated helix trajectory tracks. The equation system
is solved using a robust accelerated GPU brute-force algorithm based
on interval arithmetic. All code is written using the Julia 
programming language.
\end{abstract}




\end{frontmatter}



\section{ Introduction}

\label{sec:Introduction}
The problem studied in this paper is the identification of a helix from a set of drift distance hits given by a Full Stereo Cylindrical Drift Chamber (FSCDC). The search for a helix in a drift chamber with a noisy or non-noisy data set has a long story in particle physics \citep{Frühwirth2020}. 

The idea of applying the Apollonius problem  \citep{Apollonius}
to the search for tracks in an axial drift chamber is not new \citep{Alicke}. This problem is used as  guideline to find 
one equation which satisfy the drift distance hits  produced by helix trajectory in a full stereo drift chamber. However, to the best authors' knowledge, this is the first time that the problem of Apollonius is generalised to a FSCDC by taking into account the stereo angle of the wires. 

Using a classical root-finding solver in order to recover the helix parameters by solving the system of equations deduced from the Apollonius’ problem and applying it  to each subset of five hits has some disadvantages like the non-convergence of the calculation without a good initial estimation of the solution and also the total computing time increases  exponentially by exhaustively checking all subsets. 

The method proposed here is robust using GPU-accelerated brute force algorithm. It uses interval arithmetic, and defined from the parameter space, a multidimensional interval space and a vote accumulator. The dimension of the space parameter is equal to the number of parameters to be determined. Each cell (which corresponds to  a multidimensional interval) of the vote accumulator contains the sum of the votes of all hits. The solution is in the cells where the maximum of votes in the accumulator are found. The last part is similar to that carried out in a Hough transform  \citep{Frühwirth2020,Alicke}. Here, the total computing time is proportional to the number of hits. 

For this study, the drift distances are generated from a homemade toy model coming from the helix trajectories of charged particles passing through a FSCDC. The toy model and the method are fully written using the Julia programming language \citep{bezanson2017julia}. This language provides both high-level programming and high-performance. The applicability of using the Julia language for high energy physics research is also explored in \citep{Eschle_2023}.

\section{Drift distance equation based on the Apollonius' problem}

\subsection{Case of axial drift chamber}
With a  wire of an Axial Cylindrical Drift Chamber (ACDC), it's possible to define a circle consisting of the drift distance as the radius of the circle and the point of intersection of the wire with an orthogonal plane to the axis of the ACDC as the center of the circle \citep{Frühwirth2020}. As the wire is parallel to the axis of the ACDC, the point of intersection is always the same whatever the position of the plane  transverse to  the axis of the chamber. 

Considering 3 circles defined by outputs from 3 axial wires and according to the Apollonius' problem \citep{Apollonius}, only up to 8 circles are tangent to these 3 circles.  One of these 8 circles is the base circle of the helix trajectory projected on the transverse plane assuming a uniform magnetic field parallel to the axis of the ACDC.
In order to obtain the correct circle solution, we need to consider more other signal outputs \citep{Alicke}.

\subsection{Case of a full stereo drift chamber}
The method for the axial case cannot be used without modification for the stereo case. As the wire is no longer parallel to the axis of the FSCDC (due to the stereo angle), the circle tangent to the base circle of the trajectory cannot be defined directly from the measurement unless the hit
position on the wire is known which is not the case here. Thus, given a stereo wire, our objective is to define in a given transverse plane, the virtual drift distance in order to be able to use the Apollonius' problem to recover the helix trajectory parameters.

The coordinates $(x,y,z)$ of a helix point at the transverse curvilinear abscissa  $s_{ \perp } $ are given by the equations (\ref{HelixEq_a}) to  (\ref{HelixEq_d})~:

{\setlength\arraycolsep{2pt}
\begin{subequations}
  \begin{align}
      &x  = x_c +  R \sin{ \left( \rho s_{ \perp } +\varphi_0 \right)} , \label{HelixEq_a} \\
      &y = y_c -  R \cos{ \left( \rho s_{ \perp } +\varphi_0 \right)} , \label{HelixEq_b} \\
       &z  = z_0 + \lambda  s_{ \perp }, \label{HelixEq_c}\\
      &\varphi_0 = {\rm{atan2}}{\left( x_{ \rm ref.} - x_c , y_c - y_{\rm ref.}\right)}, \label{HelixEq_d}
\end{align}
\end{subequations}}

where $(x_c , y_c )$ and  $R = \frac{1}{\rho}$ are respectively the center coordinates and the radius of the base circle of the helix defined in a plane  transverse  to the axis of the FSCDC,  $z_0$  the Z-coordinate of one helix point and  $\lambda $  the ratio of the longitudinal component to the transverse component of the track momentum. The angle $\varphi_0$ is  the azimuthal angle  of the tangent to the base circle at the distance of closest approach to one reference point with coordinates $(x_{ \rm ref.}, y_{\rm ref.})$. More information on the definitions used here can be found in~\citep{Avery}.

Given a stereo wire numbered $i$ defined by the stereo angle $ \tau_i$, the intersection coordinates $ x_i^{  \wireAx},  y_i^{  \wireAx}$  of the stereo wire in a chosen transverse plane and the wire projection angle  $\phi_{ s i}$ in this plane,  the signed drift distance  $ d_{ i}^{  \st }$ to this wire $i$ satisfies the equation (\ref{ApoEQ3D}) (see \ref{appendA})~: 

{\setlength\arraycolsep{2pt}
\begin{align}
          &\left(x_i^{  \wireAx } - x_c  \right)^2+\left(y_i^{  \wireAx }
            - y_c \right)^2  -\left( R +  d_i^{ \wireAx }  \right)^2     = 0, 
            \label{ApoEQ3D}
\end{align}}

where the expression of $ d_i^{ \wireAx }$, function of the stereo wire $i$ and the helix parameters, is given in the equation  (\ref{defdiAX_b}). The absolute value of reconstructed signed drift distance $ d_i^{ \wireAx }$ can be interpreted as the radius of a circle with the center coordinates  $ ( x_i^{  \wireAx},  y_i^{  \wireAx})$ in the chosen transverse plane. This circle is tangent to base circle of the helix and therefore can be use in the Apollonius' problem  \citep{Apollonius}.

\section{Performance studies using Julia language}
\subsection{Use interval arithmetic for finding the helix parameters}
To solve the system of equations built from the 
equation (\ref{ApoEQ3D}) taking into account  all hits, the interval 
arithmetic \citep{Moore} is used  with the help of the  IntervalArithmetic.jl package \citep{Sanders}. A 4-dimensional vote accumulator based on the space of helix parameters to be determined is built.  All the parameters belong in it. The  parameter space is discretized. The momentum of the particle is fixed to its true value in order to reduce the number of voxels. Only four parameters have  to be  determined. The sign of $\lambda$ remains unknown. The multidimensional interval   $ \bar{x}_c    \times   \bar{y}_c  \times  \bar{R} \times \bar{z}_0  $ is referred here as voxel. 
Still to reduce the number of voxels considered, only voxels satisfying the condition (\ref{ApoVoteDisk}) are taken into account~:

\begin{subequations}
  \begin{align}
  &   0 \in    \bar{x}_c^2 + 
 \bar{y}_c^2  -\left( \bar{R}  +  \bar{R}_{\rm Disk}\right)^2 ,\label{ApoVoteDisk} \\ &\bar{R}_{\rm Disk} = [- 10, 10] \ \rm cm, \label{ApoVoteDisk_b}
\end{align}
\end{subequations}

where $ \bar{R}_{\rm Disk}$, given in equation  (\ref{ApoVoteDisk_b}), is the size of the transverse region where the particle is emitted.

By scanning all voxels considered, each measurement $i$ votes for a set of voxels  according to the condition (\ref{ApoVote})~:

\begin{equation}
  0 \in    \left(x_i^{  \wireAx } - \bar{x}_c  \right)^2+\left(y_i^{  \wireAx }  - \bar{y}_c \right)^2  -\left( \bar{R} + d_i^{ \wireAx }   \right)^2,  \label{ApoVote}
\end{equation}

where the interval arithmetic part is the interval arithmetic extension of the left part of the equation (\ref{ApoEQ3D}). For each measurement and for the two drift distance signs,   
   the condition (\ref{ApoVote}) is verified  using the 
   considered voxel. If zero belongs to the returned 
   interval, the 4D accumulator cell corresponding 
   to the voxel is incremented by one count. This procedure is applied for
     each sign of the  $\lambda$ value, so two 4D accumulators are used. 
     
     The expression (\ref{ApoVote}) reduces the dependency problem of the interval arithmetic \citep{Moore} by minimizing the number of times where the intervals $\bar{x}_c ,\bar{y}_c $ and $  \bar{R}  $ appear. Because of returned interval overestimation, we cannot conclude that the solution (the helix parameters to be determined) are in the voxel when the condition (\ref{ApoVote}) is true but when the condition (\ref{ApoVote}) is false the solution is not in the voxel. This property is no longer guaranteed because
     the variables appearing in the expression of $ d_i^{ \wireAx } $, given in the equation (\ref{defdiAX_b}),  are not considered as interval. Their values are taken in the middle of the voxel. This approximation reduces the computing time by about a factor of 10. 
     
\subsection{Full Stereo Drift chamber Toy Model to evaluate the performances}
In order to evaluate the performances, using the Julia language, a toy model based on the FSCDC is built like the one used by the COMET experiment \citep{COMET}. This model can be generalised to any other stereo drift chambers. The detector is immersed in a uniformed magnetic field of 1 Tesla parallel to the axis of the FSCDC (Z axis). This detector has 4,986 sensitive wires (20 layers). The absolute stereo angle of the wires relative to Z axis is in the order of 100 mrad. The inner and the outer radius of the detector are respectively of the order of 496  mm and 837 mm, its length is the order of 1,536 mm.

For this study, electrons are generated with a momentum of 105 MeV/c and with a transversal momentum constrained in order to have a helix with less than one turn in the FSCDC and reaching the three first layers of sensitive wires. The electrons are emitted isotropically from the origin point $(0,0,0)$ which is located at the middle of the FSCDC symmetry axis. The drift distance is defined as the smallest distance between the wire and the helix. Due to the space between wires, the maximum drift distance is taken to 8 mm. A hit is defined by this drift distance and the coordinates of the ends of the wire. Typically, the range of number of hits is from 25 to 62 with a mean value equal to about 45 hits.

\subsection{Results and discussions}
Massive GPU parallelism is used by expressing voting operations as a broadcast operation on an array of voxels that is executed efficiently on GPU hardware using the CUDA.jl package \citep{besard2018juliagpu}.
Constrained by the CPU and the GPU memory size and by the time on the computation, the total number of voxels used here is lower than 1 million which limits the voxel size.  In order to have the required precision an iterative procedure of three steps is used. 

After each iteration, the values and the errors of the  parameters are computed  using voxels corresponding to selected cells where the maximum of votes in the accumulator are found. The central values and the size of the selected voxels are used to calculate the parameter values with their errors.

After each iteration, a hit finding is performed using the interval arithmetic. An multidimensional interval is constructed. Its central value is given by the parameter values of the reconstructed helix  and its radius is given by their errors  multiplied by a factor of 5 to insure all hits are found. One hit is selected when the condition  (\ref{ApoVote}) is true for this multidimensional interval. Only the selected hits are used for the next iteration but all hits are always considered for the hit finding.

In the first step, the initial global size grid is defined by the entire space of parameter solution and the voxel size is fixed, see table (\ref{tabGridDef}). The reference point is set to $(0, 0)$. In the second step,  the reference point is taken as the position of measurement point closest to the recovered circle. The voxel size is reduced and the global size grid remains unchanged, see table (\ref{tabGridDef}). In the third step, a new reference point is defined as in the previous step. The search is performed in a space parameters  centered on the value determined by the previous step and the voxel size is reduced to obtain the final desired accuracy, see table (\ref{tabGridDef}). 

In order to test the method,  5,000 particle tracks  coming from the origin point   $(0, 0, 0)$ have been generated. The computing time is about 1.2 s for processing one particle track measurements with 4 NVIDIA K80 (12 Go). This time is divided by a factor of 6 with NVIDIA V100 (32 Go). These GPU's have been provided by the CNRS/IN2P3 Computing Center~\citep{CCIN2P3}. 

\begin{figure}
\begin{subfigure}[b]{0.46\textwidth}
\includegraphics[width=\linewidth]{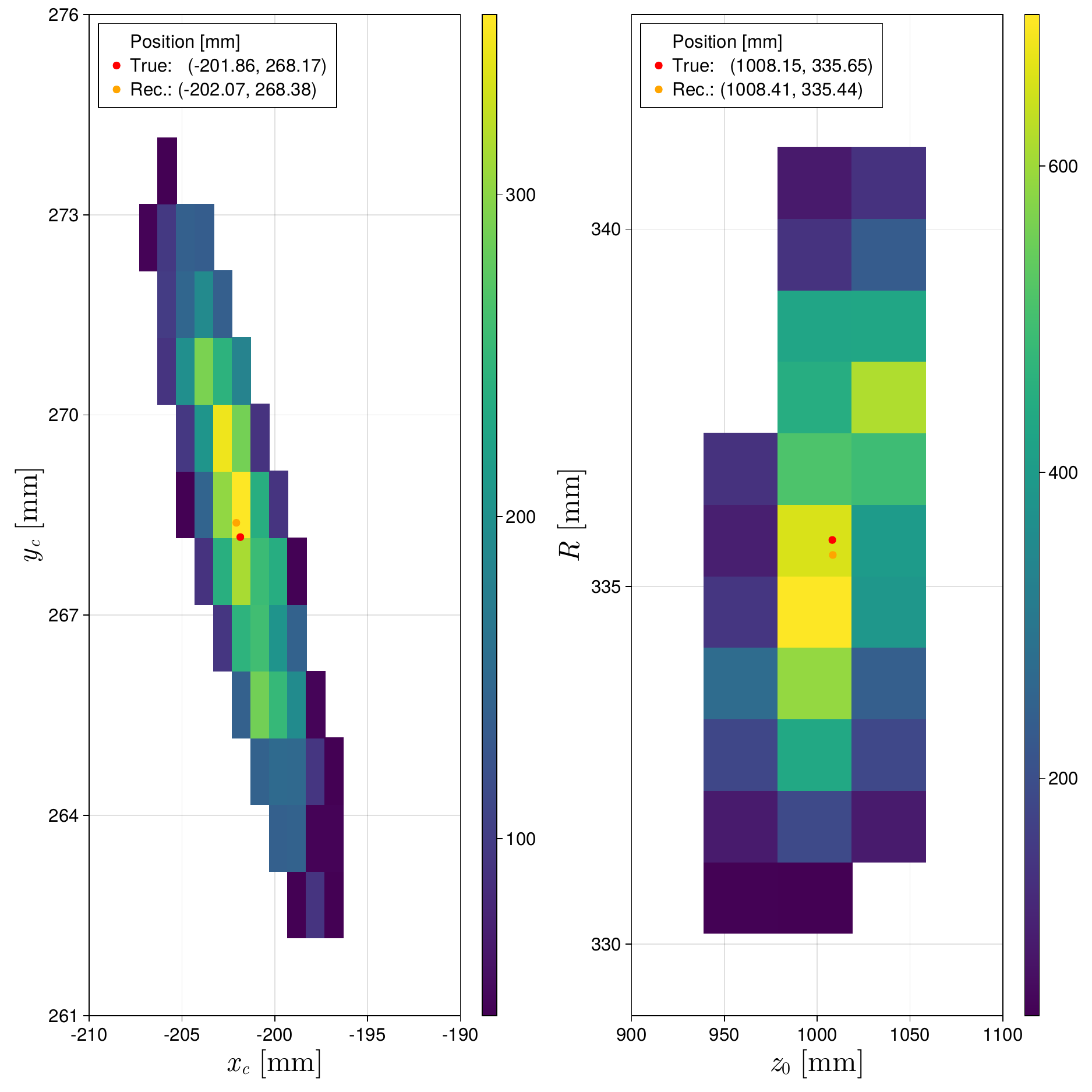}
\caption{Exemple of 2D projections  on the coordinates  $(x_c, y_c)$ (left)  and  $(z_0, R)$ (right) of the selected voxels for one event.
\label{accumulatorProjection}  }
\end{subfigure}
\begin{subfigure}[b]{0.46\textwidth}
\begin{subfigure}[b]{0.48\textwidth}
\includegraphics[width=\linewidth]{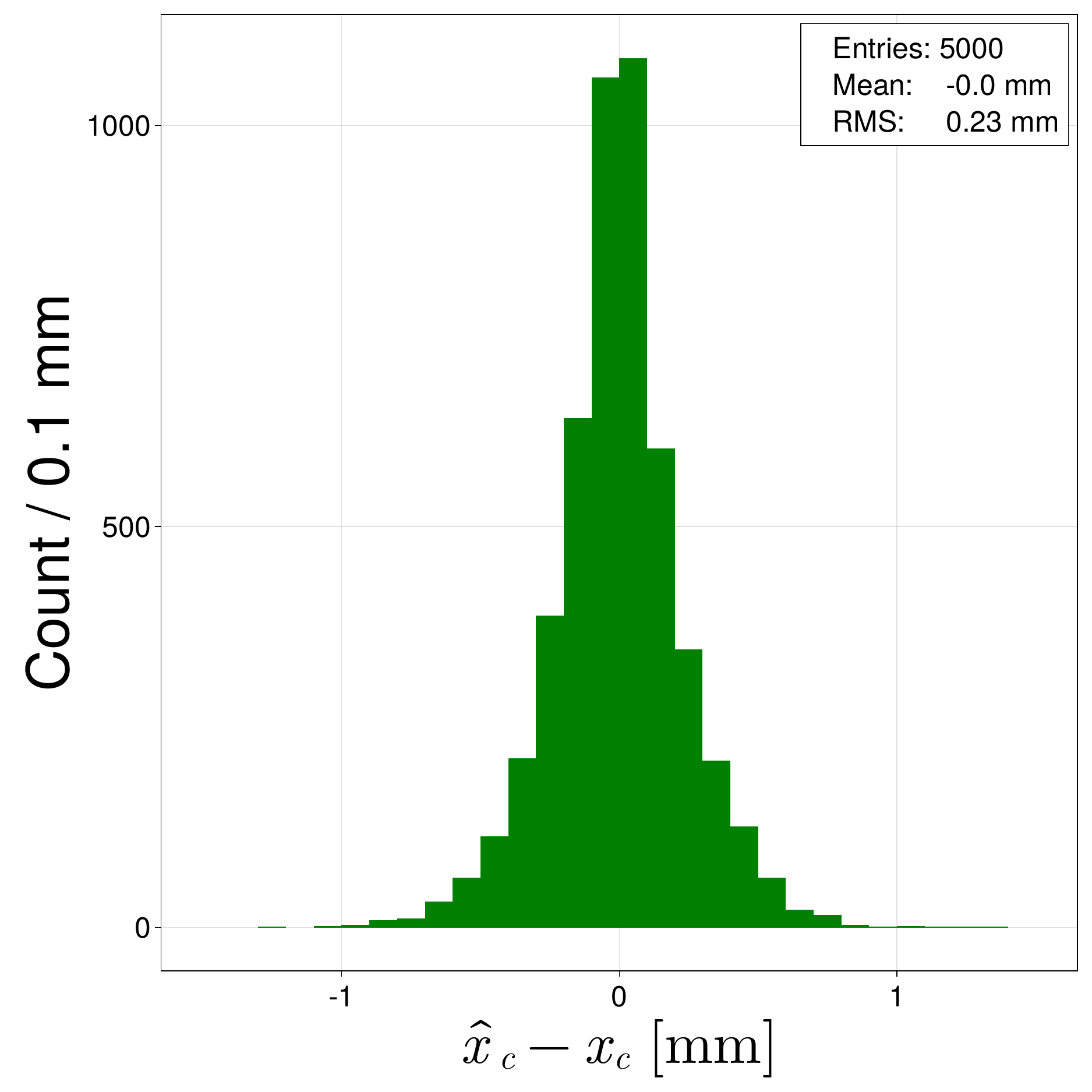}
\includegraphics[width=\linewidth]{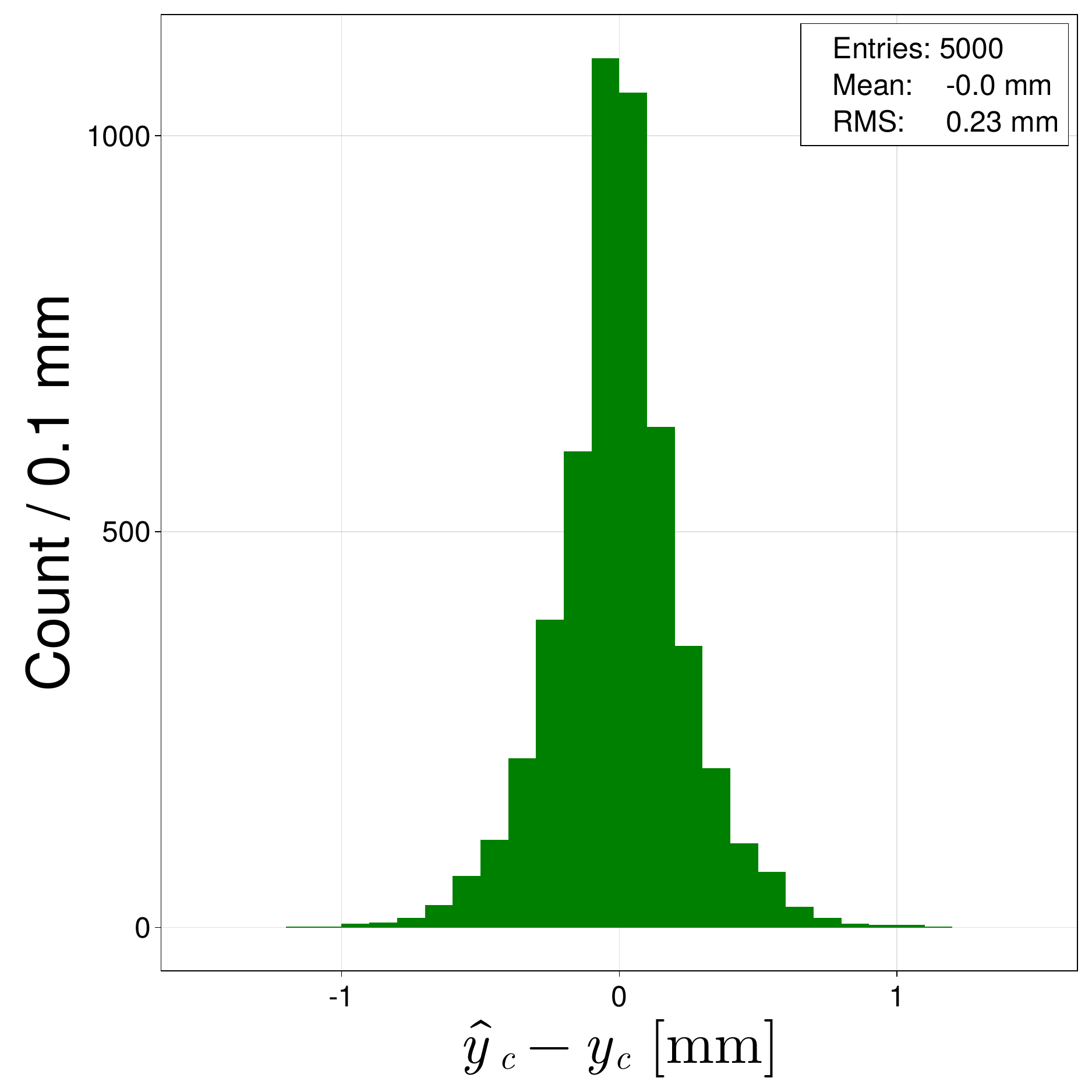}
\end{subfigure}
\begin{subfigure}[b]{0.48\textwidth}
\includegraphics[width=\linewidth]{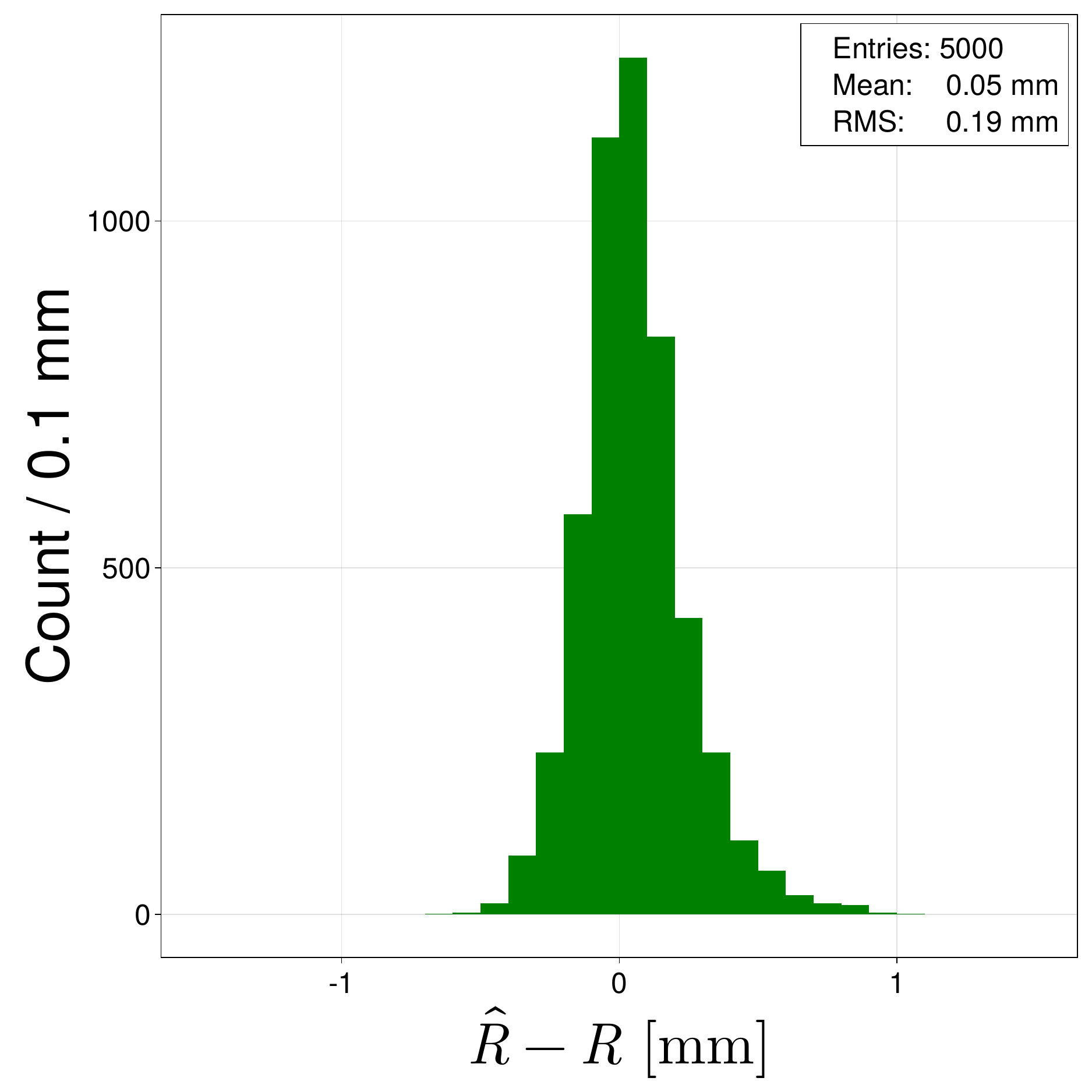}
\includegraphics[width=\linewidth]{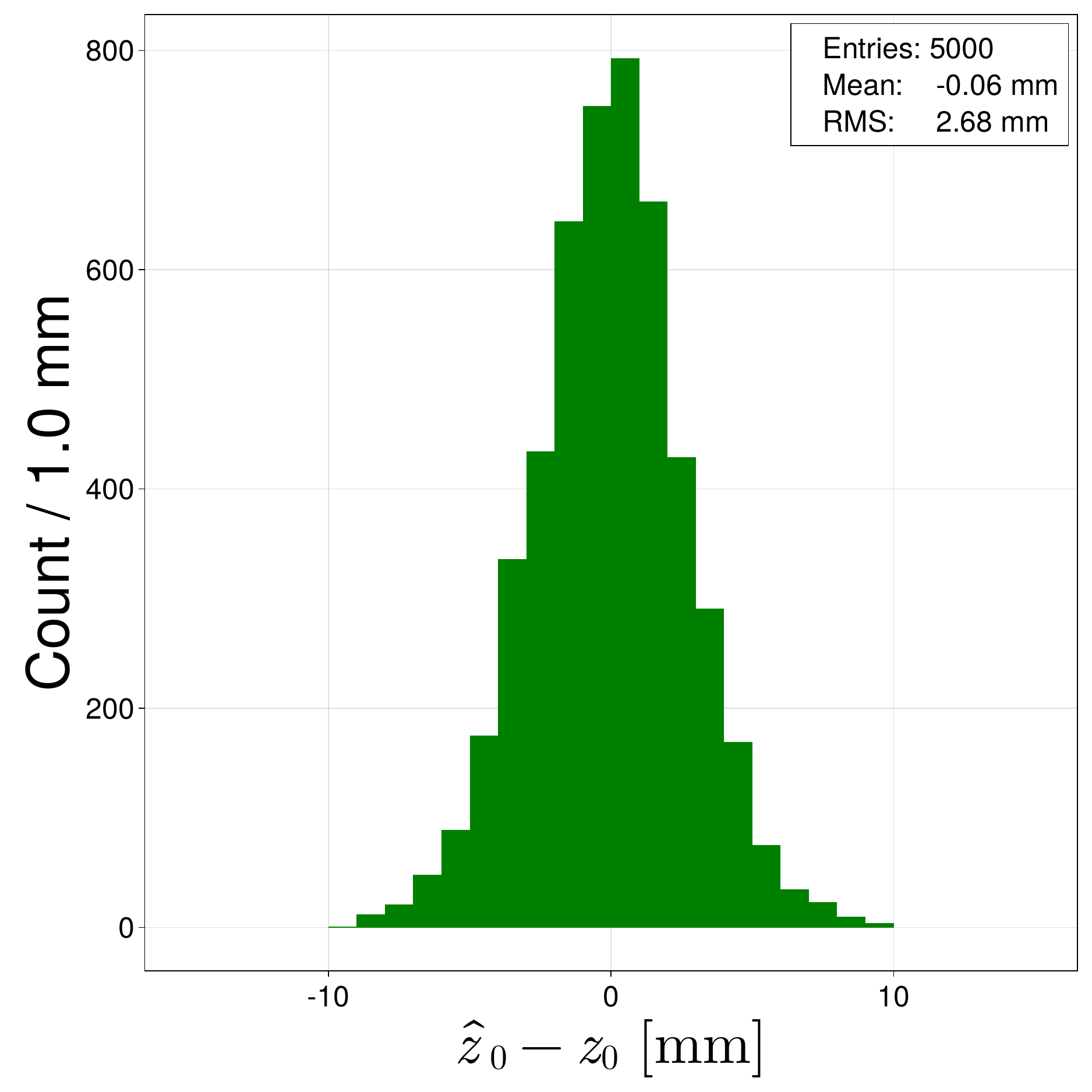}
\end{subfigure}
\caption{Distributions of the differences between reconstructed value (hat symbol) and true value (no hat symbol) for the $(x_c, y_c,z_0, R)$  helix parameters. \label{disRecPar} }
\end{subfigure}
\caption{}
\end{figure}

The figure (\ref{accumulatorProjection}) shows, after the last iteration,   the 2D projections on the coordinates  $(x_c, y_c)$  and  $(z_0, R)$ of the
selected voxels. Only cells having a number of votes strictly greater than the maximal value of the votes minus 10 are retained for display. The colorbar gives the total number of votes in the 2D projection. The figure (\ref{disRecPar}) shows, after the last iteration, the distributions of the difference between reconstructed and true value for the $(x_c, y_c,z_0, R)$  helix parameters. The figures  (\ref{accumulatorProjection}) and (\ref{disRecPar}) are plotted using Makie.jl package \citep{Makie}.

The resolution evolution of the reconstructed helix parameters and  the wrong reconstructed $\lambda$ sign fraction according to the iteration number is given in table (\ref{HelixParametes}). 
At the last iteration, the Root Mean Square (RMS) reached is the order of $200 \  \rm \mu m$
for $x_c, y_c$ and $R$ and the order of $3 \ \rm mm $ for the $z_0$. The choice of the $\lambda$  sign is achieved by comparing the maximal value of the votes  in the two 4D accumulators corresponding to each  $\lambda$  sign. The 4D accumulator retained is the one with the greatest value. According to table (\ref{HelixParametes}) there is no wrong reconstructed $\lambda$ sign at the end. 

The fraction of reconstructed parameters having an
accuracy smaller than 1 mm for the values of  $(x_c, y_c, R)$ and smaller than 10 mm for the value of $z_0$ is  99.7 \%. The results are similar when the particles do not pass through the reference point $(0,0)$ taken in the first iteration. For example, with particles emitted at the point  $(50  \ \rm mm, 0, 0)$ or $(0, 50 \ \rm mm, 0)$  this fraction becomes respectively 98.9 \% and 99.0 \%. 

The resolution is directly related to the fineness of the grids. The results obtained are really up to those expected and show that the chosen method is very well suited to solve this kind of problem.

\begin{table}
\begin{tabular}{ |p{1cm}
|p{10cm}
|p{4cm}| }
 \hline
Iter. & Global size Grid ($\rm mm^4$) & Voxel Size ($ \rm mm^4$) \\
 \hline
1  & $[-450, 450 ]_{x_c} \times [-450, 450 ]_{y_c} \times  [225, 375]_{R} \times  [0, 1500]_{z_0}$ & $18 \times 18 \times 3.75 \times 75 $ \\
 \hline
 2  & $[-450, 450 ]_{x_c} \times [-450, 450 ]_{y_c} \times  [225, 375]_{R} \times  [0, 1500]_{z_0}$ & $10 \times 10 \times 3.75 \times 75 $ \\
 \hline
 3  & $ [\hat{x}_c \pm 25   ] \times   [\hat{y}_c \pm 25  ] \times [\hat{R} \pm 25  ] \times [\hat{z}_0 \pm 400 ]   $ &  $1 \times 1 \times 1 \times 40 $ \\
 \hline
\end{tabular}
\caption{ Grid definition.  $\hat{x}_c, \hat{y}_c, \hat{R} $  and $  \hat{z}_0$ are the  reconstructed values of helix parameters.\label{tabGridDef}   }
\end{table}

\begin{table}
\begin{tabular}{ |p{1cm}| 
|p{1cm}|p{1cm}|
|p{1cm}|p{1cm}|
|p{1cm}|p{1cm}|
|p{1.2cm}|p{1cm}|
|p{1.1cm}|            }
 \hline
  & \multicolumn{2}{|c||}{ $ x_c$ (mm)} &  \multicolumn{2}{|c||}{ $y_c$ (mm)}  & \multicolumn{2}{|c||}{ $R$ (mm)} & \multicolumn{2}{|c||}{$z_0$ (mm)} & $\lambda$ sign \\
 \hline
 Iter.   & Mean & RMS & Mean  & RMS &Mean & RMS & Mean & RMS &  WRSF   \\
 \hline
1 & 0.04 & 9.08  & 0.15 & 8.93 & 5.05 & 9.39  & -86.53 & 252.17 & 16.6\% \\
 \hline
 2  & 0.01 & 3.81 & 0.07 & 3.78 & 2.03 & 3.43 & -3.02 &45.51 & 4.7 \% \\
 \hline
 3 & 0.0 & 0.23 & 0.0 & 0.23 & 0.05 & 0.19 & -0.06 & 2.68 & 0 \% \\
 \hline
\end{tabular}
\caption{Resolution of   $x_c, y_c,R$ and $z_0$  reconstructed helix parameters and Wrong Reconstructed $\lambda$ Sign Fraction (WRSF) 
 obtained with a statistic of 5,000 events. \label{HelixParametes}}
\end{table}

\section{Conclusion}
  A new efficient tracking and hit finding method using the Apollonius' problem of circles as guideline  in combination with the interval arithmetic  has been developed,  fully written using the Julia programming language and implemented on GPU. This method is successfully applied to a set of signal hits in a stereo drift chamber using Monte Carlo simulated data. The procedure is also highly configurable and can be easily adapted to multiple environments. We intend to continue this work  by studying the background hit rejection power of this method, by integrating this algorithm into the COMET experiment software and extend this method to the case of quasi-homogeneous magnetic field.

\section*{Acknowledgements}
We gratefully acknowledge support from the CNRS/IN2P3 Computing Center (Lyon - France) for providing computing and data-processing resources needed for this work.



\appendix

\section{Stereo drift distance equation  \label{appendA}}
The goal of this section is to prove the equation (\ref{ApoEQ3D}) using the Apollonius’ problem as guideline. This equation gives in a reference plane  transverse to the axis of the FSCDC the condition for a circle defined from a measurement point and the parameters of the helix to be tangent to the projection of  the base circle of the helix on this reference plane. This reference plane is orthogonal to the helix axis. The magnetic field is uniform and parallel to the axis of the FSCDC. This section is largely inspired by the paper  \citep{Avery}. 

A stereo wire numbered $i$ can be defined by the stereo angle $ \tau_i$, the intersection coordinates  $ x_i^{  \wireAx}\left( 0\right),  y_i^{  \wireAx}\left( 0\right)$ of the stereo wire in the transverse plane defined at $ z_i^{\rm wire} = 0 $  and  the wire projection angle  $\phi_{ s i}$ in this plane.  The coordinates $  \left( x_i^{ \rm St.},  y_i^{ \rm St.}, z_i^{\rm wire}\right) $  of a point of this stereo wire  are given in the equations (\ref{wirePar_a}) and (\ref{wirePar_b})~:

{\setlength\arraycolsep{2pt}
\begin{subequations}
  \begin{align}
      &    x_i^{ \rm St.} =  x_i^{ \rm \wireAx }\left( 0\right) + z_i^{\rm wire} \sin{\phi_{ s i}}  \tan{\tau_i}, \label{wirePar_a}   \\
      &  y_i^{ \rm St.}   =   y_i^{ \rm \wireAx }\left( 0\right) -  z_i^{\rm wire}  \cos{\phi_{ s i}}\tan{\tau_i}. \label{wirePar_b}
\end{align}
\end{subequations}}

The signed drift distance   $d_{ i}^{\st}$ satisfy the equation (\ref{defSignedDriftDist})~:

\begin{equation}
    \overrightarrow{OM_i^{ \rm St.}} = \overrightarrow{OM_i} - d_{ i}^{  \st } \frac{\vec{u}_{d i} }{u_{d i}}, 
   \label{defSignedDriftDist}
   \end{equation}

where the  $M_i^{ \rm St.}$  wire point    and  the  $M_i$  helix point  are closest to each other. The vector $\vec{u}_{d i }$   is parallel  to the drift line and its norm $u_{d i} $ is given in  the equation  (\ref{normudi})~:

{\setlength\arraycolsep{2pt}
\begin{subequations}
  \begin{align}
   u_{d i} &= \sqrt{1+ 2 \lambda  \sin{ \Phi_i }   \tan{\tau_i}  + \lambda^2  \tan^2{\tau_i}  + \cos^2{ \Phi_i }  \tan^2{\tau_i} },  \label{normudi} \\  
    \Phi_i &=  \rho   s_{ \perp i} +  \varphi_0 -\phi_{ s i},
\end{align}
\end{subequations}}

and $ s_{ \perp i}$ is the transverse curvilinear abscissa at the  $M_i$ point.

By using the signed drift distance definition given in equation (\ref{defSignedDriftDist}) and the helix definition given in the equations  (\ref{HelixEq_a}) to  (\ref{HelixEq_d}), the equations  (\ref{EQSinter_a}) to (\ref{EQSinter_e}) are derived~:

{\setlength\arraycolsep{2pt}
\begin{subequations}
  \begin{align}
    & x_i^{  \wireAx} - x_c   + z_{i }^{ \rm St.}  \sin{\phi_{ s i}} \tan{\tau_i} -   \lambda  \frac{ d_i^{  \st }  }{u_{d i}}  \cos{\phi_{ s i}}  \tan{\tau_i} =   \left( R +  \frac{ d_i^{  \st } }{  u_{d i}}    \right) \sin{ \left( \Phi_i + \phi_{ s i} \right) },  \label{EQSinter_a}  \\
   &      y_i^{  \wireAx} - y_c   - z_{i }^{ \rm St.}  \cos{\phi_{ s i}} \tan{\tau_i} -   \lambda  \frac{ d_i^{  \st }  }{u_{d i}}  \sin{\phi_{ s i}}  \tan{\tau_i} = -\left( R +  \frac{ d_i^{  \st } }{  u_{d i}}    \right) \cos{ \left( \Phi_i + \phi_{ s i} \right) },  \label{EQSinter_b}\\ 
     & z_{i }^{ \rm St.} =   z_0 + \lambda  s_{ \perp i} -  z_{ \rm ref.}   - \frac{ d_i^{  \st }  }{u_{d i}}   \cos{ \Phi_i } \tan{\tau_i},  \label{EQSinter_c}\\
    & x_i^{ \rm \wireAx } =  x_i^{ \rm \wireAx }\left( 0\right) + z_{ \rm ref.} \sin{\phi_{ s i}}  \tan{\tau_i},  \label{EQSinter_d}   \\
    & y_i^{ \rm \wireAx } =  y_i^{ \rm \wireAx }\left( 0\right) -  z_{ \rm ref.} \cos{\phi_{ s i}}\tan{\tau_i},  \label{EQSinter_e}
\end{align}
\end{subequations}}

where $z_{ \rm ref.} $ is an arbitrary constant parameter which will be determined later. 

Now after a straight forward computation the circle equation  (\ref{defdiAX_a})  is obtained~:
 
{\setlength\arraycolsep{2pt}
\begin{subequations}
  \begin{align}
       &  \left(x_i^{  \wireAx } - x_c  \right)^2 + \left(y_i^{  \wireAx } - y_c \right)^2   -\left( R +    d_i^{  \wireAx }   \right)^2    = 0, \label{defdiAX_a}\\         
              &  d_i^{  \wireAx } = \sqrt{  \left( R +  \frac{ d_i^{  \st } }{  u_{d i}}    \right)^2  - 2  c_i^{\tau}   \tan{\tau_i} - \left(  z_{i }^{ \rm St.}  \tan{\tau_i}    \right)^2  - \left(   \lambda  \frac{   d_i ^{  \st } }{  u_{d i}}    \tan{\tau_i}   \right)^2  } - R, \label{defdiAX_b}
\end{align}
\end{subequations}}

 as well as the angular equation (\ref{CSAXtoPHI_a})~:
 
{\setlength\arraycolsep{2pt}
 \begin{align}
 &\Phi_i  =   \Phi_i^{\ax} +  {\rm{atan2}}\left( s_i^{\tau}  \tan{\tau_i},  \left(   R_i^{  \wireAx } \right)^2   + c_i^{\tau}  \tan{\tau_i}     \right),  \label{CSAXtoPHI_a}
\end{align}}

where

{\setlength\arraycolsep{2pt}
\begin{subequations}
 \begin{align}
    &R_i^{  \wireAx } = \sqrt{\left(x_i^{  \wireAx } - x_c  \right)^2+\left(y_i^{  \wireAx } - y_c \right)^2  }, \\
    &c_i^{\tau} = -   \lambda   \frac{ d_i^{ \st}  }{ u_{d i}  } s_i^{\ax} +  z_{i }^{ \st} c_i^{\ax}, \\
    &   s_i^{\tau} = -   \lambda   \frac{ d_i^{ \st}  }{ u_{d i}  } c_i^{\ax} -  z_{i }^{ \st} s_i^{\ax}, \\
  &c_i^{\ax}  =  - \left( x_c  -  x_i^{  \wireAx } \right) \sin \phi_{ s i} + \left( y_c -  y_i^{  \wireAx } \right) \cos \phi_{ s i}, \\
 & s_i^{\ax}  =  - \left( x_c  -  x_i^{  \wireAx } \right) \cos \phi_{ s i} - \left( y_c -  y_i^{  \wireAx } \right) \sin \phi_{ s i}, \\
&\varphi_i^{  \wireAx } = {\rm{atan2}}\left( x_i^{  \wireAx } - x_c ,    y_c - y_i^{  \wireAx }    \right), \\
&\Phi_i^{\ax} = \varphi_i^{  \wireAx } -  \phi_{ s i}.
     \label{defEQApo1}
\end{align}
\end{subequations}}

The absolute value of reconstructed signed drift distance $ d_i^{ \wireAx }$ given in the equation  (\ref{defdiAX_b}) can be interpreted as the radius of a circle with the center coordinates  $ ( x_i^{  \wireAx},  y_i^{  \wireAx})$ in the reference transverse plane defined at $  z_i^{\rm wire}  = z_{ \rm ref.}$ . The equation  (\ref{defdiAX_a}) shows that this
circle is tangent to base circle of the helix and therefore can be use in the Apollonius' problem  \citep{Apollonius}.

The tangent of the stereo angle  is small so the approximate solution of the equation (\ref{CSAXtoPHI_a}) is given in the equation (\ref{ApproxPhi}) as a second-degree polynomial in $\tan{\tau_i}$~:

{\setlength\arraycolsep{2pt}
\begin{subequations}
  \begin{align}
    &\Phi_i \approx  \Phi_i^{\ax} +  \Phi_i^{\left(1 \right)} \tan{\tau_i} +  \Phi_i^{\left(2 \right)}  \tan^2{\tau_i},  \label{ApproxPhi} \\
     &  \Phi_i^{\left(1 \right)} =  - \frac{ z_{i }^{\ax}  \sin{  \Phi_i^{\ax} }   + \lambda d_i^{  \st } \cos{  \Phi_i^{\ax} } }{R_i^{  \wireAx }},  \label{ApproxPhi_b} \\
      \begin{split}
       &   \Phi_i^{\left(2 \right)} = \frac{1}{2 (R_i^{  \wireAx })^2} \left[  \lambda R   z_{i }^{\ax}    - \lambda   z_{i }^{\ax}  \left(  R - 2  d_i^{  \st } \right)    \cos{ \left(  2 \Phi_i^{\ax} \right)   }   \right.    \\
       & \quad \quad \quad  \quad \quad \quad \quad  \left.   +     \left(  (z_{i }^{\ax})^2  +  d_i^{  \st } \left( R_i^{  \wireAx }  + \lambda^2 \left(   R +  R_i^{  \wireAx } - d_i^{  \st } \right)      \right)    \right)      \sin{ \left(  2 \Phi_i^{\ax} \right)   }       \right], 
  \end{split}  \label{ApproxPhi_c} \\       
          &  z_{i }^{\ax}   =   z_0 - z_{ \rm ref.}  + \lambda R \left( \varphi_i^{\ax}  - \varphi_0 \right).  \label{ApproxPhi_d}
\end{align} 
\end{subequations}}

The corrections  $\Phi_i^{(1)}$ and $ \Phi_i^{(2) }$, given in  the equations (\ref{ApproxPhi_b}) to (\ref{ApproxPhi_d}),  are deduced using Mathematica \citep{Mathematica} and  the $\Phi_i^{(1)}$ value  has been  checked by hand calculation. The choice  of $ z_{ \rm ref.} = z_0$ reduces the size of these corrections. All numerical results are in good agreement whether in taking $ z_{ \rm ref.} = 0$ or solving numerically the equation (\ref{CSAXtoPHI_a}).

\bibliographystyle{elsarticle-num-names} 
\bibliography{cas-refs}





\end{document}